\documentclass{emulateapj}

\usepackage{apjfonts}
\usepackage{amsmath}
\usepackage{amssymb}
\usepackage{natbib}
\usepackage{xspace}
\usepackage{graphicx}
\usepackage{rotating} 
\usepackage{subfigure}
\usepackage{float}

\newcommand{\e}[1]{\ensuremath{^{#1}}\xspace}

\newcommand{\Msun}{\ensuremath{M_\odot}\xspace}

\newcommand{\nh}{$N_{\text{H}}$\xspace}

\newcommand{\pexrav}{\textsc{pexrav}\xspace}
\newcommand{\xspec}{\textsc{xspec}\xspace}

\newcommand{\etal}{et al.\xspace}
\newcommand{\ka}{K$\alpha$\xspace}

\newcommand{\chidof}{$\chi^{2}/$dof\xspace}
\newcommand{\dchidof}{$\Delta \chi^{2}$/dof}
\newcommand{\nustar}{\textsl{NuSTAR}\xspace}

\newcommand{\xmm}{\textsl{XMM-Newton}\xspace}
\newcommand{\chandra}{\textsl{Chandra}\xspace}
\newcommand{\fluxunits}{erg\,cm$^{-2}$\,s$^{-1}$\xspace}
\newcommand{\colunits}{$\times 10^{22}$ cm$^{-2}$\xspace}

\shorttitle{Absorption Variability in NGC 1365}
\shortauthors{Rivers et al.}

\begin{document}

\title{The Multi-Layer Variable Absorbers in NGC 1365 Revealed by \xmm and \nustar} 
\author{E.~Rivers\altaffilmark{1}, G.~Risaliti\altaffilmark{2,3}, D.J.~Walton\altaffilmark{4,1}, F.~Harrison\altaffilmark{1},  P.\ Ar\'evalo\altaffilmark{5}, F.E. Baur\altaffilmark{6,7}, S.E. Boggs\altaffilmark{8}, L.W.\ Brenneman\altaffilmark{3}, M. Brightman\altaffilmark{1},  F.E. Christensen\altaffilmark{9}, W.W. Craig\altaffilmark{8}, F.\ F\"urst\altaffilmark{1}, C.J. Hailey\altaffilmark{10}, R.C.\ Hickox\altaffilmark{11}, A.\ Marinucci\altaffilmark{12}, J. Reeves\altaffilmark{13}, D.~Stern\altaffilmark{4}, W.W. Zhang\altaffilmark{14}}

\altaffiltext{1}{Cahill Center for Astronomy and Astrophysics, California Institute of Technology, Pasadena, CA 91125, USA} 
\altaffiltext{2}{INAF -- Osservatorio Astrofisico di Arcetri, Largo Enrico Fermi 5, 50125 Firenze, Italy} 
\altaffiltext{3}{Harvard-Smithsonian Center for Astrophysics, 60 Garden Street, Cambridge, MA 02138, USA} 
\altaffiltext{4}{Jet Propulsion Laboratory, California Institute of Technology, Pasadena, CA 91109, USA} 
\altaffiltext{5}{Instituto de F\'isica y Astronom\'ia, Facultad de Ciencias, Universidad de Valpara\'iso, Gran Bretana N 1111, Playa Ancha, Valpara\'iso, Chile} 
\altaffiltext{6}{Pontificia Universidad Cat\'olica de Chile, Instituto de Astrof\'isica, Casilla 306, Santiago 22, Chile}
\altaffiltext{7}{Space Science Institute, 4750 Walnut Street, Suite 205, Boulder, Colorado 80301, USA} 
\altaffiltext{8}{Space Sciences Laboratory, University of California, Berkeley, CA 94720, USA}
\altaffiltext{9}{DTU Space, National Space Institute, Technical University of Denmark, Elektrovej 327, DK-2800 Lyngby, Denmark} 
\altaffiltext{10}{Columbia Astrophysics Laboratory, Columbia University, New York, NY 10027, USA} 
\altaffiltext{11}{Department of Physics and Astronomy, Dartmouth College, 6127 Wilder Laboratory, Hanover, NH 03755, USA} 
\altaffiltext{12}{Dipartimento di Matematica e Fisica, Universita degli Studi Roma Tre, via della Vasca Navale 84, 00146 Roma, Italy} 
\altaffiltext{13}{Astrophysics Group, School of Physical and Geographical Sciences, Keele University, Keele, Staffordshire, ST5 5BG, United Kingdom}
\altaffiltext{14}{NASA Goddard Space Flight Center, Greenbelt, MD 20771, USA} 


\begin{abstract}

Between July 2012 and February 2013, \nustar and \xmm performed four long-look joint observations of the type 1.8 Seyfert, NGC 1365.
We have analyzed the variable absorption seen in these observations in order to characterize the geometry of the absorbing material.
Two of the observations caught NGC 1365 in an unusually low absorption state, revealing complexity in the multi-layer absorber which 
had previously been hidden.  We find the need for three distinct zones of neutral absorption in addition to the two zones of ionized 
absorption and the Compton-thick torus previously seen in this source.
The most prominent absorber is likely associated with broad line region clouds with column densities of around $\sim\,$10$^{23}$ cm$^{-2}$ 
and a highly clumpy nature as evidenced by an occultation event in February 2013.
We also find evidence of a patchy absorber with a variable column around $\sim\,10^{22}$ cm$^{-2}$ and a line of sight covering fraction of 0.3--0.9 
which responds directly to the intrinsic source flux, possibly due to a wind geometry. 
A full-covering, constant absorber with a low column density of $\sim\,1 \times$ 10$^{22}$ cm$^{-2}$ is also present, though the location of this low density haze is unknown.

\end{abstract}

\keywords{Galaxies: active -- X-rays: spectra -- Galaxies: Individual: NGC 1365}

\section{Introduction}


\begin{deluxetable*}{lcccc}
   \tablecaption{Observation Details \label{tabobs}}
   \tablecolumns{5}
   \startdata
\hline
\hline\\[-1mm]
Observation  				&	1	&	2	&	3	&	4	\\[1mm]
\hline\\[-1mm]
\nustar ObsID				& 60002046002/3 & 60002046005& 60002046007& 60002046009 \\[1mm]
FPMA Net Exposure Time (ks)	& 	77	&	66	&	74	&	70	\\[1mm]
FPMB Net Exposure Time (ks)	& 	77	&	66	&	74	&	70	\\[1mm]
\hline \\[-1mm]
\xmm ObsID				& 0692840201 & 0692840301 & 0692840401 & 0692840501 \\[1mm]
MOS Net Exposure Time (ks)	& 	134 	& 122	& 105	& 122  \\[1mm]
PN Net Exposure Time (ks)	&	118	&	108	&	93	&	116 \\[-1mm]
\enddata
\end{deluxetable*}

NGC~1365 is a Seyfert 1.8 active galactic nucleus (AGN) that exhibits a highly complex and variable X-ray spectrum.
In addition to the power law continuum which is thought to arise in a hot corona very close to the central supermassive black hole,
the X-ray spectrum shows a strong Compton reflection hump peaking at 20--30~keV (see e.g., George \& Fabian 1991, Walton \etal 2010) and a 
prominent Fe K emission complex (Risaliti \etal 2009, Risaliti \etal 2013), both of which are signatures of reflection off Compton-thick material.
Extended X-ray emission from plasma and starburst activity below $\sim\,$2~keV have been characterized by \chandra (Wang \etal 2009) 
and \xmm (Guainazzi \etal 2009).  Absorption lines from ionized Fe species in the $\sim\,$7--8~keV range are thought to arise in a 
highly ionized ($\xi \sim 10^3 - 10^4$\, erg\, cm\, s$^{-1}$) high-velocity outflow ($v \sim 1000-5000$ km s$^{-1}$) 
that varies on timescales of days to months (Risaliti \etal 2005; Brenneman \etal 2013).
The mass of the central black hole has been estimated from the $H\beta$ width to be $M_{\rm BH} \,\sim\,2 \times 10^{6}$ \Msun (Risaliti \etal 2009 and references therein),
implying an Eddington ratio of 0.02 to 0.12 $L/L_{\rm Edd}$.

Due to the complexity of this source, broad energy coverage is necessary in order to accurately characterize the X-ray spectrum.
Four simultaneous \xmm and \nustar observations were taken in 2012 and 2013 with the primary aim of studying the broad Fe \ka line and prominent reflection hump 
in this source likely associated with relativistic reflection from the inner regions of the accretion disk (Risaliti \etal 2013).  
Measurements of the black hole spin from modeling the relativistic reflection indicate a rapidly rotating black hole with a dimensionless spin parameter $a \gtrsim 0.95$
(Risaliti \etal 2013; Walton \etal 2014), consistent with previous measurements (Brenneman \etal 2013), though with higher signal-to-noise broadband spectra.
Other work on this dataset by Kara \etal (2014) showed that the Fe \ka line and Compton reflection hump lag the continuum on 
the same timescales and Walton \etal (2014) found a correlation between the strengths of the Fe \ka line and the Compton hump.  This connection is 
important in establishing that the broadness in the line is indeed associated with reflection from the accretion disk, rather than being due to absorption or some other spectral variance.

These joint  \xmm/\nustar  observations revealed strong variability in the flux as well as in the spectral shape.  
Walton \etal (2014) found that the majority of the spectral variability was due to variable line-of-sight absorption, 
a result that was subsequently confirmed by principle component analysis (Parker \etal 2014).
Two of the observations caught the source in an unusually low absorption state, and the third of these observations actually showed an almost complete uncovering of the source.  
Braito \etal (2014) have analyzed the \xmm RGS spectrum from this observation and found evidence for a mildly ionized wind absorber in 
addition to the previously known highly ionized wind absorber that is only evident when the source is uncovered and at a high luminosity level.
This paper aims to analyze the variable absorption seen in all four joint \xmm/\nustar observations in order to further characterize the geometry of the absorbing material in this source.

Variable absorption is not uncommon in Seyfert galaxies on a wide range of timescales, from hours (MCG--6-30-15: Marinucci \etal 2014) to months 
(NGC~3227 -- Lamer \etal 2003; Cen~A -- Rivers \etal 2011), to years (NGC~3516 -- Turner \etal 2008; NGC~2110 -- Rivers \etal 2014).
Changes over a timescale of years may be due to a global change in the amount of material surrounding the supermassive black hole, 
while changes on shorter timescales (hours to months) are likely a result of inhomogeneous material.  
Lamer \etal (2003) characterized an occultation event in NGC~3227 that showed a smooth rise and fall in the column density due to the 
ingress into and egress out of the line of sight of a clump of material which was thought to be part of the broad line region (BLR).
A similar event lasting around 60 days was seen in Cen~A (Rivers \etal 2011) but was found to originate in the infrared torus, 
consistent with the clumpy torus models set forth by Nenkova \etal (2008).

In the past, the absorption column in the line of sight to NGC~1365 has been seen to exhibit rapid variability on timescales of hours to days 
(Risaliti \etal 2009).  This is thought to be due to BLR clouds passing through the line of sight (Maiolino \etal 2010).  
On longer timescales, the absorbing column has been seen to vary widely with a range of column densities spanning
\nh$\,\sim\, 10^{22} -10^{24}$ cm$^{-2}$ (Connolly \etal 2014).  Connolly \etal (2014) also noticed an anti-correlation between the column 
density and the intrinsic luminosity which they suggest could be explained by winds with variable launch radii.

This paper is structured in the following way: Section 2 contains details of the observations and data reduction, Section 3 describes the spectral analysis
with limited interpretation, and Section 4 discusses our results and the conclusions we can draw from them.



\begin{figure*}
   \includegraphics[trim=5cm 0cm 5cm 0cm,clip,height=\textwidth,angle=270]{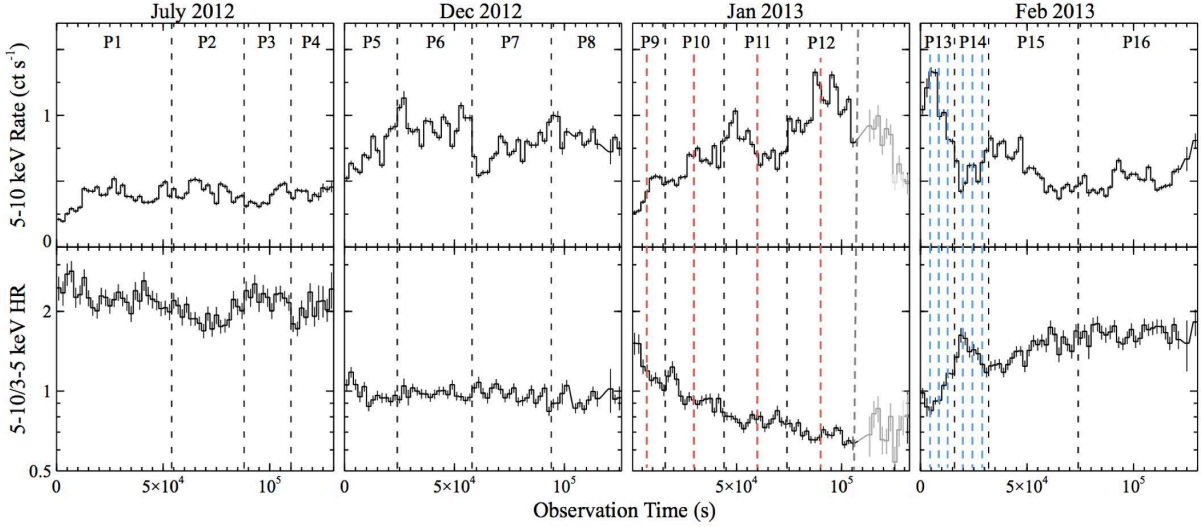}
  \caption{Light curves for all four observations showing 5--10~keV observed flux (top) and 5--10/3--5~keV hardness ratio (bottom).  Dashed lines indicate time-resolved analysis intervals: P1--16 (black), the 8 subintervals of observation 3 (red) and the 8 subintervals of observation 4 (blue).  Note that the gray data points at the end of observation 3 were excluded from analysis.  A background flare occurred during this time creating large uncertainties, particularly for time-resolved analysis.}
  \label{figlc}
\end{figure*}


\begin{figure*}
   \plottwo{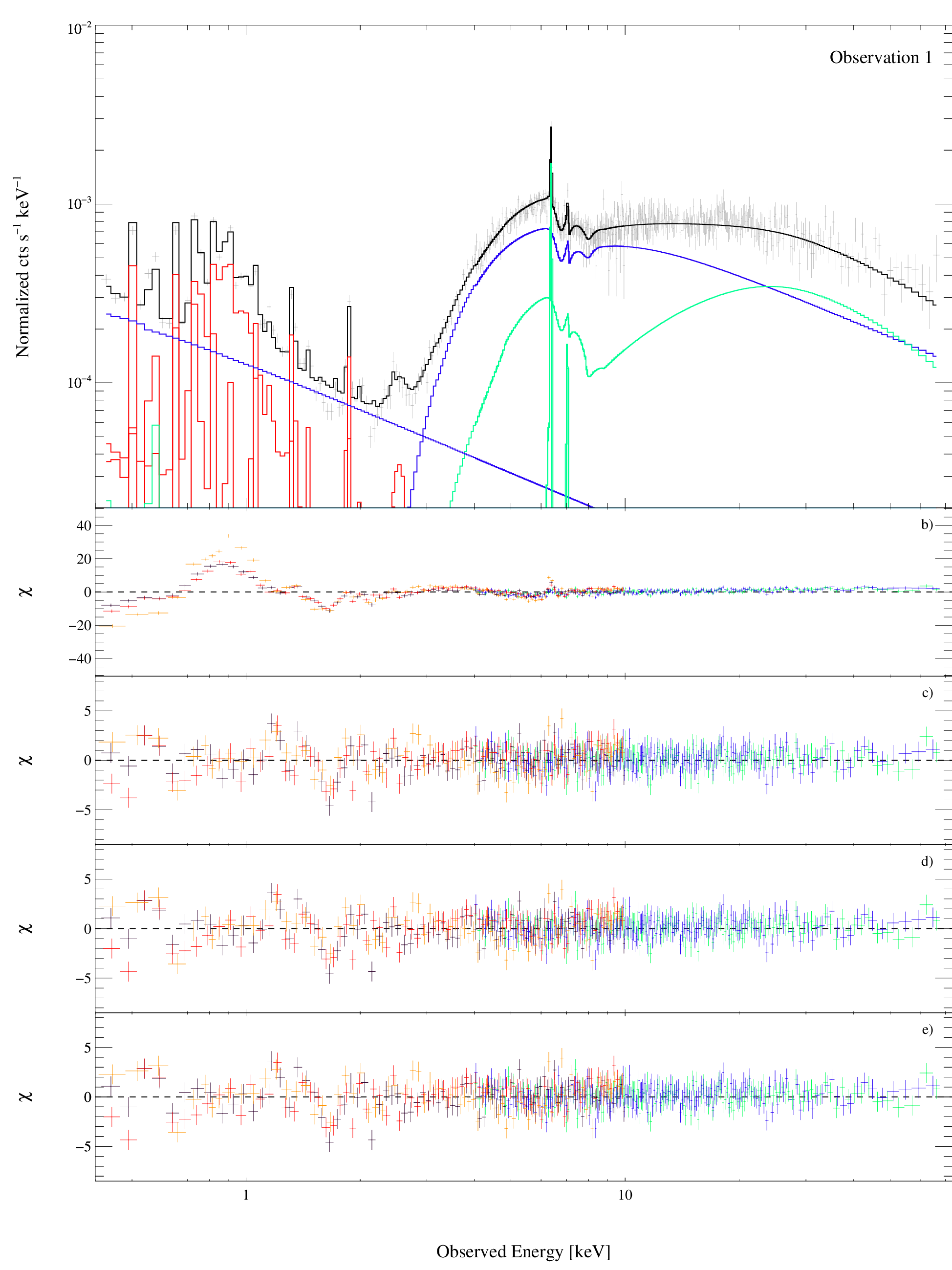}{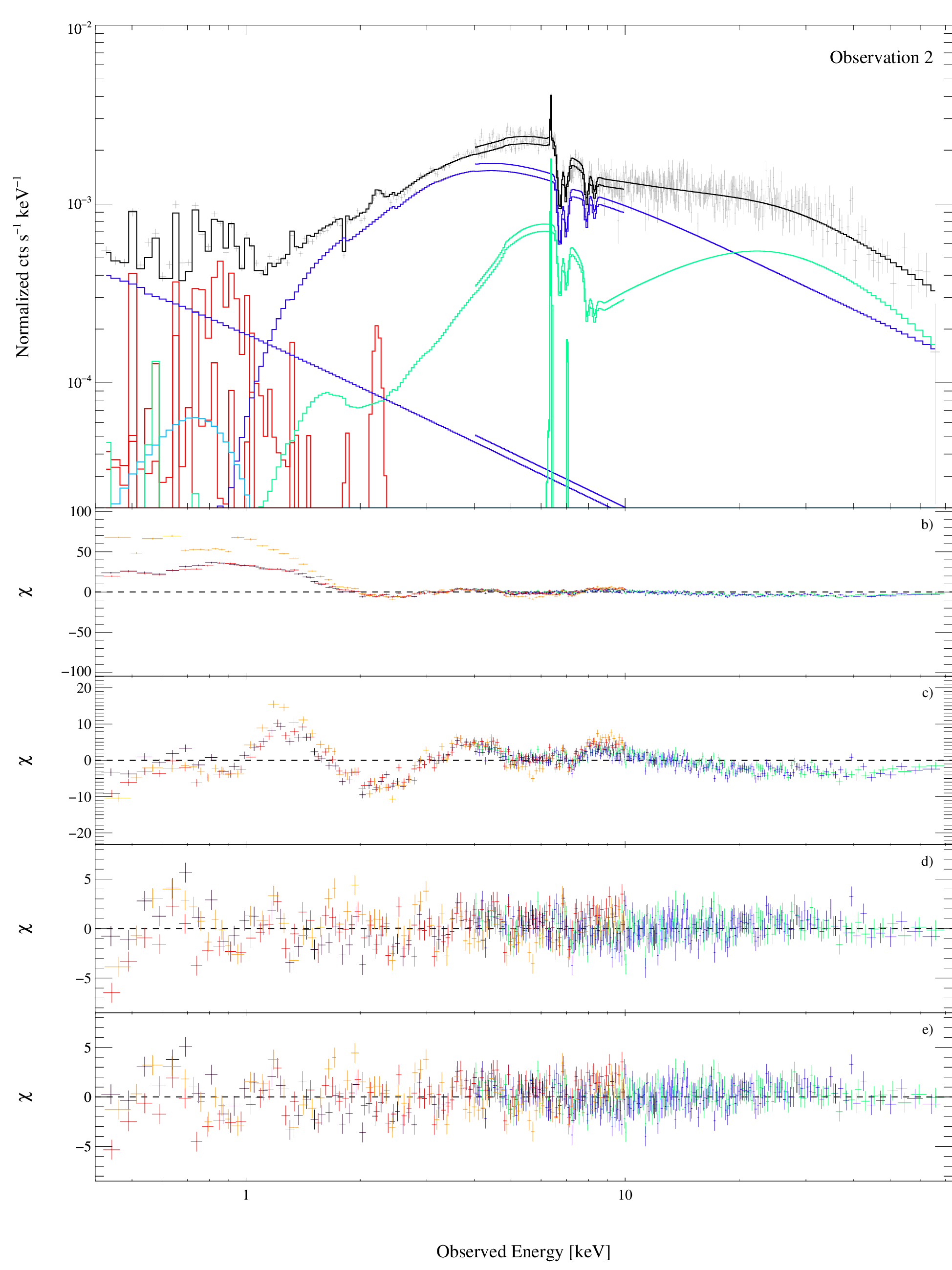}
   \plottwo{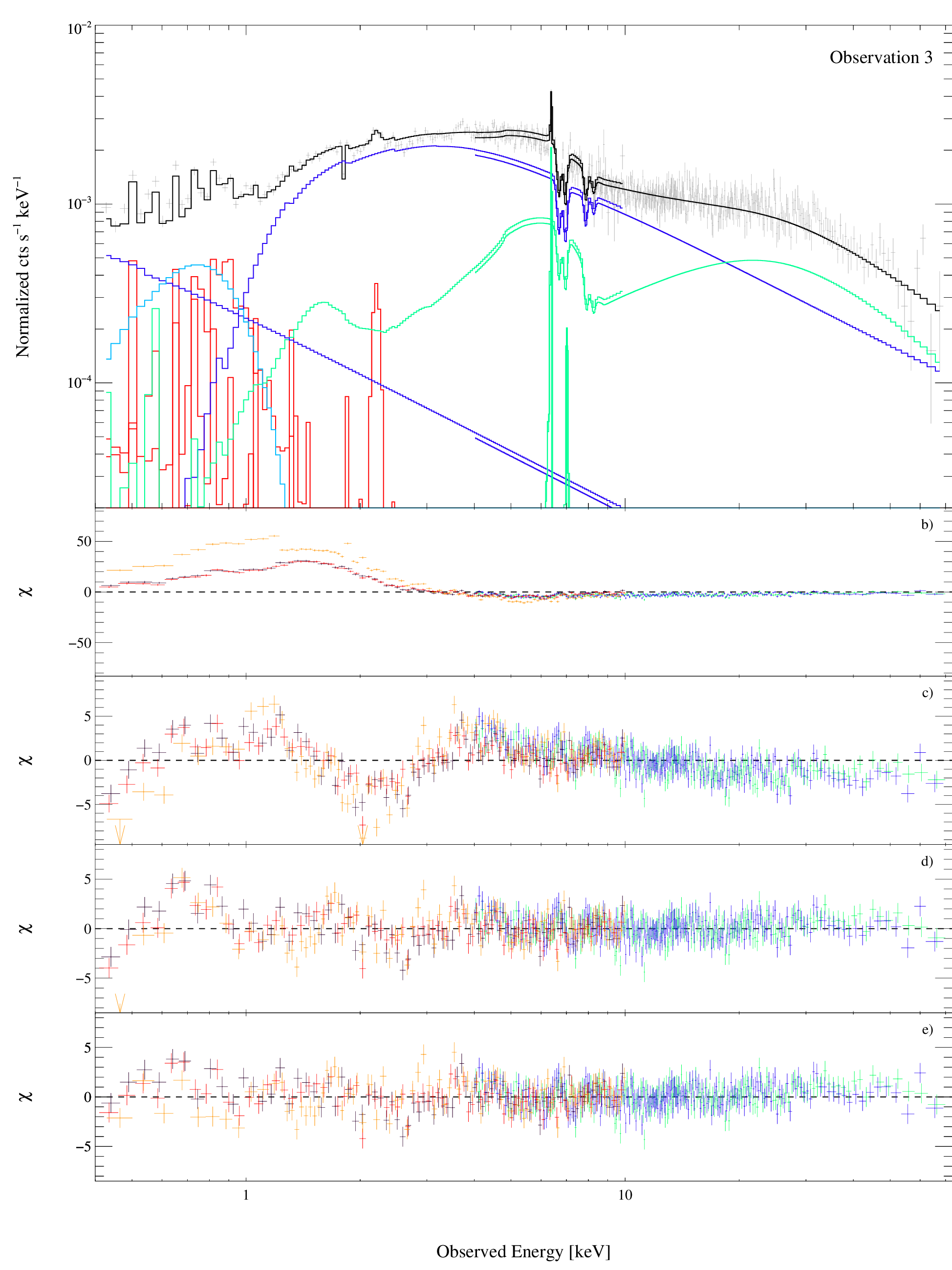}{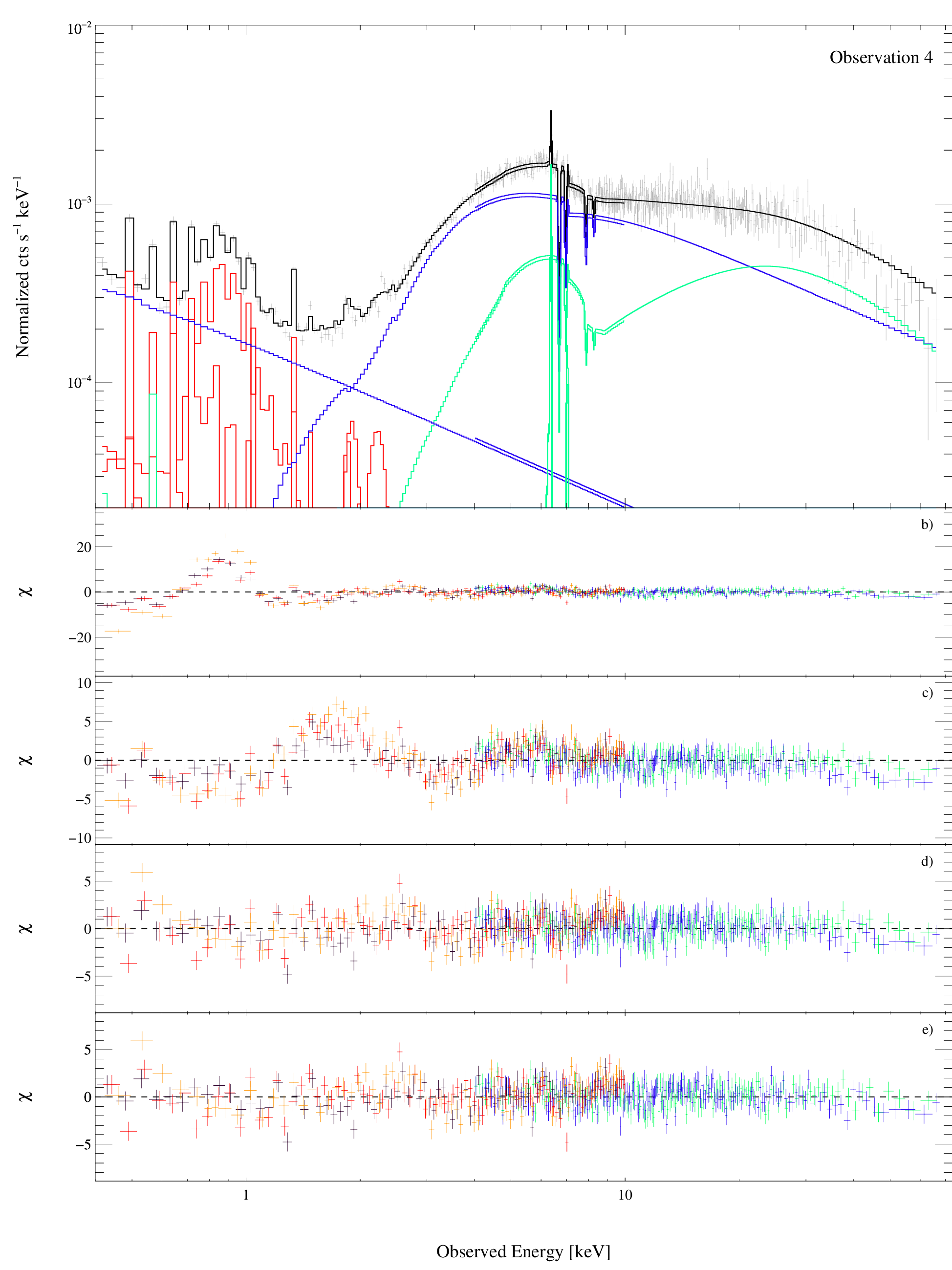}
 \caption{Panel (a) shows the 0.4-70~keV unfolded spectra of all four observations (PN + FPMA in gray) with full model (black), absorbed and scattered power laws (blue), extended plasma components (red), relativistic and cold/distant reflection (green), and the phenomenological Gaussian component (light blue).  Residuals are shown in panels b-e with red and orange data points corresponding to \xmm MOS1 and MOS2, respectively, purple data points corresponding to PN data, and blue and green data points corresponding to \nustar FPMA and FPMB, respectively.  Panel (b) shows the best fit to the initial model as described in Section 3.1 before including the extended plasma, scattered power law or additional absorption complexity.  Panel (c) shows the best fit to a model including the extended plasma but with only one neutral absorber component (partial-covering).  Panel (d) shows the best fit to a model without the phenomenological Gaussian at 0.68~keV.  Panel (e) shows the best fit to our final model for each observation as described in the text with parameters given in Table 2. }
  \label{figspecall}
\end{figure*}



\section{Observations and Data Reduction}\label{sec:analysis}

Data were taken 2012 July, 2012 December, 2013 January, and 2013 February with \xmm and \nustar simultaneously.  
Table \ref{tabobs} shows a log of the observations.  All extractions were done using HEASOFT v.6.13.

In order to explore the changing spectral parameters, we have performed time-resolved spectral analysis.  
In addition to characterizing the four individual observations, each has been further subdivided into a total of 
16 intervals (four per observation) based on flux and hardness level as seen in Figure~\ref{figlc}, identical to those of Walton \etal (2014).
Additionally, we have performed analysis on very short timescales where rapid changes in the absorption column
are observed: eight half-intervals for observation 3 and eight 4~ks intervals for the first 32~ks (two intervals) of observation 4.

\subsection{\xmm Reduction}

We reduced the \xmm data for all four observations using the Science Analysis System (SAS v13.0.0) following the procedure detailed in the online guide and Walton \etal (2014).
We processed the data using EPPROC and EMPROC for the EPIC-pn (Str\"uder \etal 2001) and EPIC-MOS (Turner \etal 2001) data, respectively.
Spectra and light curves were extracted from circular source and background regions with 40\arcsec\ and 50\arcsec\ radius for the pn and MOS, respectively.
Response and ancillary response matrices were generated using the FTOOLs RMFGEN and ARFGEN.

During the last portion of observation 3 the source reached a peak in the flux level sufficient for pile-up to be a concern. 
We found some evidence for pileup in the MOS below 10~keV, and therefore excluded the central 
8\arcsec\ for the pn and 10\arcsec\ for the MOS for observation 3 and interval P12 (the last interval of observation 3).
Note that we did not include data from observation 3 past 110~ks 
where a background flare occurred, possibly contaminating the data (see Figure \ref{figlc}).

\subsection{\nustar Reduction}

We reduced data from both \nustar modules, FPMA and FPMB (Harrison \etal 2013), using the standard pipeline in the \nustar Data Analysis Software v1.1.1.
Instrumental responses were taken from the \nustar CALDB v20130315.
The unfiltered event files were cleaned with the standard depth correction, which significantly reduces the internal 
background at high energies, and SAA passages were excluded from our analysis. 
For both modules we extracted spectra and light curves from a 100\arcsec\ circular source region and a 100\arcsec\ background region on the same chip as the source.
We grouped the \nustar spectra with a minimum of 25 counts per bin.




\begin{figure}
   \plotone{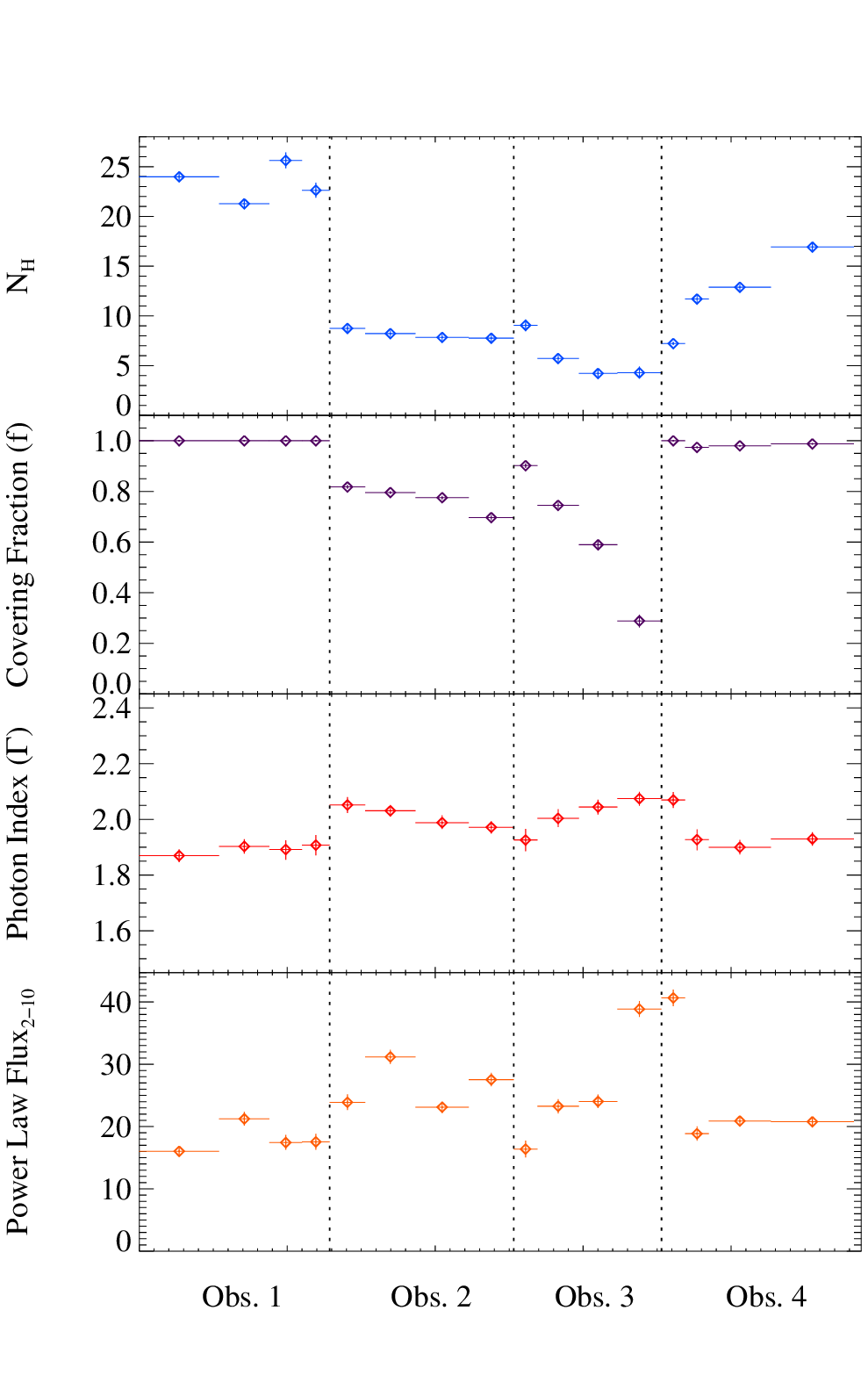}
  \caption{Parameter evolution from time-resolved fitting the 16 intervals (four per observation): column density, covering fraction, photon index and the unabsorbed 2--10~keV flux..  The variation in column density is consistent with the hardness ratio evolution seen in Figure~\ref{figlc}.  The photon index is stable over each observation and does not vary greatly between observations while the intrinsic flux varies greatly.  \nh is in units of 10$^{22}$ cm\e{-2} and power law flux is in units of 10\e{-12} \fluxunits.}
  \label{figpars}
\end{figure}


\begin{deluxetable*}{l@{\hspace{2mm}}ccc@{\hspace{2mm}}c@{\hspace{2mm}}c@{\hspace{2mm}}c@{\hspace{2mm}}c@{\hspace{2mm}}c@{\hspace{2mm}}c@{\hspace{2mm}}r}
   \tablecaption{Broadband Model Parameters \label{tabpar}}
   \tablecolumns{11}
   \startdata
\hline
\hline\\[-1mm]
Interval  & Unabsorbed & Photon  & PC Column  &  Covering  & FC Column  &  Scattered   & Soft Gaussian & Relativistic  & Distant  &  $\chi^2$/dof \\[1mm]
& Continuum  & Index & Density (\nh)  & Fraction  & Density & Power Law &  at 0.68~keV & Reflection  & Reflection & \\[1mm]
&F$_{2-10}$\tablenotemark{A} & ($\Gamma$) &  [$10^{22}$\,cm$^{-2}$] & ($f$) &  [$10^{22}$\,cm$^{-2}$] & F$_{0.5-2}$\tablenotemark{A}/F$_{2-10}$\tablenotemark{A}   & Norm ($10^{-5}$) & Norm  ($10^{-6}$)   &  Norm  ($10^{-6}$)   & \\[1mm]
\hline\\[-1mm]

Obs. 1  &   16.8$\,\pm\,$  0.3   &  1.84$\,\pm\,$0.02  &    22.4$\,\pm\,$0.2  &  1.00$^{\dagger}$ ( $> 0.99$)   &  1.0$^{*}$  &  0.27\,/\,0.38$\,\pm\,$0.01  &  -   &  1.0$\,\pm\,$0.0  &  3.0$\,\pm\,$0.2  &  1627/1243\\[1.5mm]
Obs. 2  &   27.3$\,\pm\,$  0.6   &  2.01$\,\pm\,$0.01  &     8.4$\,\pm\,$0.2  &   0.76$\,\pm\,$ 0.01  &  1.4$\,\pm\,$0.1  		&  0.44\,/\,0.52$\,\pm\,$0.03  &   4.3$\,\pm\,$ 1.3  &  2.0$\,\pm\,$0.1  &  3.4$\,\pm\,$0.4  &  2131/1353\\[1.5mm]
Obs. 3  &   23.0$\,\pm\,$  0.5   &  2.04$\,\pm\,$0.01  &     5.8$\,\pm\,$0.4  &   0.49$\,\pm\,$ 0.01  &  1.1$\,\pm\,$0.1  		&  0.78\,/\,0.75$\,\pm\,$0.05  &  20.5$\,\pm\,$ 2.4  &  2.1$\,\pm\,$0.1  &  3.2$\,\pm\,$0.5  &  1877/1347\\[1mm]
Obs. 4  &   21.6$\,\pm\,$  0.5   &  1.92$\,\pm\,$0.01  &    11.3$\,\pm\,$0.2  &   0.97$\,\pm\,$ 0.01  &  1.0$\,\pm\,$0.2  		&  0.35\,/\,0.46$\,\pm\,$0.02  &    0.2$^{\dagger}$ ($<1.0$)  &  1.5$\,\pm\,$0.1  &  3.3$\,\pm\,$0.3  &  1888/1323\\[1mm]

\hline\\[-1mm]

P1  &   16.0$\,\pm\,$  0.7   &  1.87$\,\pm\,$0.02  &    24.0$\,\pm\,$0.5  &   1.00$^{\dagger}$ ( $> 0.99$)  &  1.0$^{*}$  &  0.26\,/\,0.37$\,\pm\,$0.01  &  -   &  1.1$\,\pm\,$0.1  &  2.4$\,\pm\,$0.4  &   953/804\\[1mm]
P2  &   21.2$\,\pm\,$  1.1   &  1.90$\,\pm\,$0.03  &    21.3$\,\pm\,$0.5  &   1.00$^{\dagger}$ ( $>0.99$)  &  1.0$^{*}$  &  0.34\,/\,0.46$\,\pm\,$0.02  &  -   &  1.3$\,\pm\,$0.2  &  3.5$\,\pm\,$0.6  &   918/708\\[1mm]
P3  &   17.4$\,\pm\,$  1.2   &  1.89$\,\pm\,$0.03  &    25.6$\,\pm\,$0.8  &   1.00$^{\dagger}$ ( $>0.99$)  &  1.0$^{*}$  &  0.26\,/\,0.36$\,\pm\,$0.02  &  -   &  1.2$\,\pm\,$0.2  &  2.5$\,\pm\,$0.6  &   641/544\\[1mm]
P4  &   17.5$\,\pm\,$  1.3   &  1.91$\,\pm\,$0.04  &    22.6$\,\pm\,$0.7  &   1.00$^{\dagger}$ ( $>0.99$) &  1.0$^{*}$  &  0.26\,/\,0.35$\,\pm\,$0.02  &  -   &  1.2$\,\pm\,$0.2  &  2.6$\,\pm\,$0.7  &   569/515\\[1mm]
P5  &   23.9$\,\pm\,$  1.2   &  2.05$\,\pm\,$0.03  &     8.8$\,\pm\,$0.4  &   0.82$\,\pm\,$ 0.01  &  1.4$^{*}$  &  0.39\,/\,0.48$\,\pm\,$0.04  &   1.5$\,\pm\,$ 0.6  &  2.1$\,\pm\,$0.2  &  4.7$\,\pm\,$0.8  &   806/690\\[1mm]
P6  &   31.2$\,\pm\,$  1.1   &  2.03$\,\pm\,$0.02  &     8.2$\,\pm\,$0.3  &   0.80$\,\pm\,$ 0.01  &  1.4$^{*}$  &  0.36\,/\,0.51$\,\pm\,$0.03  &   3.1$\,\pm\,$ 0.5  &  2.6$\,\pm\,$0.2  &  5.7$\,\pm\,$0.8  &  1180/906\\[1mm]
P7  &   23.1$\,\pm\,$  0.9   &  1.99$\,\pm\,$0.02  &     7.8$\,\pm\,$0.3  &   0.78$\,\pm\,$ 0.01  &  1.4$^{*}$  &  0.39\,/\,0.54$\,\pm\,$0.03  &   1.2$\,\pm\,$ 0.5  &  1.9$\,\pm\,$0.2  &  4.5$\,\pm\,$0.7  &  1100/883\\[1mm]
P8  &   27.5$\,\pm\,$  1.1   &  1.97$\,\pm\,$0.02  &     7.8$\,\pm\,$0.4  &   0.70$\,\pm\,$ 0.01  &  1.4$^{*}$  &  0.45\,/\,0.62$\,\pm\,$0.04  &   2.8$\,\pm\,$ 0.6  &  1.7$\,\pm\,$0.2  &  3.7$\,\pm\,$0.8  &  1067/873\\[1mm]
P9  &   16.4$\,\pm\,$  1.3   &  1.93$\,\pm\,$0.04  &     9.1$\,\pm\,$0.6  &   0.90$\,\pm\,$ 0.01  &  1.1$^{*}$  &  0.49\,/\,0.69$\,\pm\,$0.08  &   1.3$\,\pm\,$ 1.7  &  1.8$\,\pm\,$0.2  &  4.9$\,\pm\,$1.0  &   529/536\\[1mm]
P10  &   23.3$\,\pm\,$  1.1   &  2.00$\,\pm\,$0.03  &     5.7$\,\pm\,$0.4  &   0.75$\,\pm\,$ 0.01  &  1.1$^{*}$  &  0.54\,/\,0.61$\,\pm\,$0.06  &  10.7$\,\pm\,$ 2.0  &  2.3$\,\pm\,$0.3  &  3.8$\,\pm\,$0.9  &  1102/774\\[1mm]
P11  &   24.0$\,\pm\,$  1.1   &  2.04$\,\pm\,$0.03  &     4.2$\,\pm\,$0.4  &   0.59$\,\pm\,$ 0.01  &  1.1$^{*}$  &  0.58\,/\,0.56$\,\pm\,$0.06  &  15.1$\,\pm\,$ 2.2  &  2.2$\,\pm\,$0.2  &  6.6$\,\pm\,$0.9  &  1102/768\\[1mm]
P12  &   38.9$\,\pm\,$  1.3   &  2.07$\,\pm\,$0.03  &     4.3$\,\pm\,$0.6  &   0.29$\,\pm\,$ 0.03  &  1.1$^{*}$  &  1.00\,/\,1.08$\,\pm\,$0.10  &  31.5$\,\pm\,$ 3.7  &  2.6$\,\pm\,$0.3  &  4.3$\,\pm\,$1.3  &  1011/874\\[1mm]
P13  &   40.7$\,\pm\,$  1.3   &  2.07$\,\pm\,$0.03  &     7.2$\,\pm\,$0.2  &   1.00$^{\dagger}$ ($>0.99$)  &  1.0$^{*}$  &  0.37\,/\,0.39$\,\pm\,$0.03  &  0.2$^{\dagger}$ ($<1.0$)  &  2.0$\,\pm\,$0.3  &  2.5$\,\pm\,$1.2  &   908/719\\[1mm]
P14  &   18.9$\,\pm\,$  1.1   &  1.93$\,\pm\,$0.04  &    11.7$\,\pm\,$0.5  &   0.97$\,\pm\,$ 0.01  &  1.0$^{*}$  &  0.33\,/\,0.43$\,\pm\,$0.05  &  0.3$^{\dagger}$ ($<1.0$) &  1.6$\,\pm\,$0.2  &  3.9$\,\pm\,$0.9  &   556/592\\[1mm]
P15  &   20.9$\,\pm\,$  0.8   &  1.90$\,\pm\,$0.03  &    12.9$\,\pm\,$0.3  &   0.98$\,\pm\,$ 0.01  &  1.0$^{*}$  &  0.31\,/\,0.42$\,\pm\,$0.03  &   2.0$\,\pm\,$ 0.6  &  1.3$\,\pm\,$0.1  &  3.4$\,\pm\,$0.6  &   917/816\\[1mm]
P16  &   20.8$\,\pm\,$  0.8   &  1.93$\,\pm\,$0.02  &    16.9$\,\pm\,$0.4  &   0.99$\,\pm\,$ 0.01  &  1.0$^{*}$  &  0.34\,/\,0.44$\,\pm\,$0.03  &   1.0$\,\pm\,$ 0.6  &  1.4$\,\pm\,$0.1  &  2.3$\,\pm\,$0.5  &  1089/950\\[1mm]

\enddata
\tablecomments{Best fit parameters for the four observations and the 16 intervals, four per observation.  PC and FC stand for partial-covering and full-covering, respectively.  The intrinsic continuum fluxes and soft scattered power law fluxes were determined using the PEGPWRLW model in \xspec. 
"*" indicates a frozen parameter.  "$\dagger$" indicates a pegged parameter.}
\tablenotetext{A}{Flux is in units of $10^{-12}$~\fluxunits.}
\end{deluxetable*}

\section{Spectral Analysis}

All spectral fitting was done in \xspec v.12.8.0 (Arnaud 1996) using the solar abundances of Anders \& Grevessse (1989) with cross sections from Verner \etal (1996). 
Uncertainties are listed at the 90\% confidence level ($\Delta \chi^2$ = 2.71 for one interesting parameter).
We included a constant offset for each instrument as a free parameter to account for known cross calibration uncertainties
and included a Galactic absorption column of 1.34 $\times 10^{20}$ cm\e{-2} in all models (Kalberla \etal 2005).

\subsection{Initial Modeling}

Initially we fit each of the four observations separately in the 3--70~keV range with an absorbed power law.  
This fit was universally poor (\chidof = 2832/1047, 6926/1145, 4551/1137, and 3968/1113 for observations 1, 2, 3, and 4, respectively)
and showed strong residuals in the Fe K bandpass ($\sim 5-9$ keV) as well as broad residuals in the 20--30 keV range 
(see Figure 1 of Risaliti \etal 2013 and Figure 1 of Walton \etal 2014 for which the continuum and reflection modeling of these data has been done in great detail).
In order to model the apparent neutral Fe emission lines we added a neutral reflection component. 
We used the \textsc{xillver} \xspec model which includes Fe K emission lines and a Compton reflection hump from a disk geometry (Garc\'ia \& Kallman 2010).
There was significant improvement in the fit statistic for each observation (\chidof = 2106/1044, 5727/1142, 3630/1134, and 3059/1110).
Note that this model choice was not a significant improvement over a phenomenological modeling using multiple Gaussian components and a neutral 
Compton hump such as from \pexrav, but we prefer a self-consistent physical model whenever possible.

Strong negative residuals in the Fe K band remained, necessitating the addition of four absorption lines from highly ionized species of 
Fe in a high velocity outflow (Risaliti \etal 2005; Brenneman \etal 2013).
We tied the velocities and line widths of the four lines, assuming intrinsic line energies of 6.70 keV, 6.97 keV, 7.88 keV, and 8.27 keV.
This greatly improved the fit statistics (\chidof = 1341/1039, 2325/1137, 2197/1129, and 1784.1/1105), 
though broad residuals still remained near the Fe \ka line and above 10 keV.
Though there are several models available which might be used at this juncture, we have elected to use \textsc{relconv}$\times$\textsc{xillver}
to model relativistic reflection from the inner parts of an ionized accretion disk.  
Inclination angle, ionization state, Fe abundance, black hole spin and normalization were left as free parameters in our initial fitting.
This model provided a good fit to the data in all four observations, (\chidof = 1087/1035, 1445/1133, 1283/1125, and 1300/1101, for the four observations, respectively).
Parameter values, detailed justification of the model, and physical interpretations for this analysis can be found in Walton \etal (2014).


Those parameters that we expect to remain constant over the timescale of our observations were frozen at their average values:
black hole spin (0.98), disk inclination (63\degr), Fe abundance (4.7).
Note that modeling the distant reflection with a torus model can lead to a much lower measurement of the Fe abundance for the distant reflector, 
however this does not change our primary results so for simplicity we adhere to the model presented in Walton \etal (2014), tying the Fe abundance between the two reflectors.
We also froze the following parameters that showed no evidence for variability over the four observations: 
ionization of the inner disk (log$\xi$=1.8), and disk emissivity index (6.75).
Note that we do not see evidence for a high energy cutoff in this source and it was therefore not included in our final model.

\subsection{Modeling the Spectrum Below 1 keV}

In order to fully characterize the partial-covering absorption in this source, we must extend our spectral analysis down to lower energies.
It was immediately evident that observations 2 and 3, and possibly observation 4, were partially unobscured in the 1--3 keV range.
We therefore replaced the full-covering absorber in our model with a partial-covering absorber, though for observation 1 the covering fraction remained pegged at 1.
Additionally, we chose to analyze the spectrum down to 0.4~keV in order to model the soft emission from diffuse plasma in the region, so that we could be
confident that our measurements of the absorption were not influenced by this component.  

The extended plasma was previously studied in detail by Wang \etal (2009) using \chandra gratings data and by Guainazzi \etal (2009) using \xmm RGS data.
These analyses determined that the plasma was likely a combination of thermal and photoionized plasma.
When the AGN is in an absorbed state the extended plasma dominates the soft X-ray flux and it is expected to remain essentially constant due to its spatial extent. 
For our modeling we used a phenomenological double \textsc{apec} component (i.e., two-temperature collisionally ionized gas) with temperatures of 0.3 and $\sim$0.8~keV, and five additional 
Gaussian components to model emission line complexes from photoionized gas at 0.50, 0.85, 1.03, 1.24, and 2.74~keV, similar to Brenneman \etal (2013).  
These components are shown in red in Figure \ref{figspecall}.  We determined the normalizations of these components using data from observation 1 only 
since it is the only observation that is fully covered and therefore allowed for the best determination of the plasma parameters.  
Since we do not expect the extended plasma to undergo any changes over the months between our observations we froze these parameters to those measured in observation 1.
We included an additional soft power law component in the modeling to account for differences in flux level below 2 keV for the four observations.
The photon index was tied to that of the primary continuum power law and in all fits was found to have a normalization around $\sim\,$2\% that of the primary power law, 
as expected for a scattered continuum component (e.g., Turner \etal 1997; Guainazzi \etal 2005; Eugechi \etal 2009).
Compared to the flux of the combined plasma components, $F_{0.5-2} = 3.6 \times 10^{-13}$ \fluxunits, the scattered power law 0.5--2 keV flux values were 
2.7$\times 10^{-13}$ \fluxunits, 4.4$\times 10^{-13}$ \fluxunits, 7.8$\times 10^{-13}$ \fluxunits, and 3.5$\times 10^{-13}$ \fluxunits for the four observations, respectively.

We applied this model to the other three observations and discovered the need for an additional layer of 
full-covering absorption, with an improvement in fit of \dchidof = 6000/1  for observation 2, \dchidof=1500/1 for observation 3, 
and \dchidof =1050/1 for observation 4 with a column density of \nh$\,\sim\,$ 1~\colunits for all three observations.
We will refer to this absorbing layer as the ``full-covering low column'' absorber for the remainder of the paper.
Given the consistency of this component over the last three observations it seems reasonable to assume that it is also present in observation 1, but it is completely degenerate with
the higher column density absorber which is fully covering during that observation.  We have included the component in our modeling of observation 1 with a fixed column density
of 1.0~\colunits in the hopes that this gives us a more accurate measurement of the column density of the partial-covering absorber layer during that observation.

We also found unexpected variability below 1~keV which was not fit by the changing absorption or the scattered power law.  
The unidentified excess appeared in the less absorbed observations, strongest in observations 2 and 3 while completely absent in observation 1.
None of the usual components that might account for this increase in flux below 1~keV were able to fit the data
(i.e., a blackbody or power law soft excess, changes in the fully-covering absorber, or reflection from ionized material). 
We therefore fit the excess with an additional phenomenological Gaussian at 0.68~keV with a width of $\sigma \sim$ 200 eV. 
Including this component resulted in an improvement in the fit of \dchidof= 200/3 for observation 3 and 
\dchidof=30/3 for observation 2 with a null hypothesis probability of $4 \times 10^{-4}$.  
It was not significant to include the component in observations 1 or 4, both of which result in a normalization of the component consistent with 0.
We investigate the source of this feature in Section 3.6.
Residuals to models excluding the full-covering low column absorber and the phenomenological Gaussian are shown for each observation in Figure \ref{figspecall}, panels (c) and (d), respectively.

\subsection{The Final Model and Time Resolved Fitting}

Our final model consisted of two collisional plasma components plus five Gaussian emission lines to model the extended plasma, a scattered power law, 
the phenomenological Gaussian at 0.68~keV, a full-covering neutral absorber (\nh $\sim$ 1~\colunits), a partial covering neutral absorber, 
four absorption lines from highly ionized species of Fe in a high velocity outflow, a continuum power law, relativistic disk reflection, and cold distant reflection.
The final form of the model in \xspec is: \textsc{apec[$\times$2]\,+\,zGauss[$\times$5]\,+\,scattered power law\,+Gauss\,+\,zphabs\,$\times$\,zpcfabs\,$\times$\,gauabs[$\times$4]\,$\times$\,(power law\,+\,relconv\,$\times$\,xillver)\,+\,xillver}.

For consistency checks, we fit all four observations simultaneously, exploring which parameters were consistent across the observations.
We find that the full-covering absorber was very steady over all three observations where it was measurable.
In observation 1 this layer is completely degenerate with the high column density layer of absorption.
We therefore included this layer in all fits to observation 1 with a column density frozen at 1.0~\colunits.  
Figure~\ref{figspecall} shows each of the four observations with final best fit model components in panel (a) and residuals to the best fit model in panel (e).  
Parameters are listed in Table \ref{tabpar}.

We then fit the 16 intervals independently in the 0.3--70~keV range using our final model.
In order to reduce degeneracy we froze the column density of the full-covering constant absorber to the time-averaged value 
for each observation and to 1.0~\colunits for all intervals of observation 1.
Best fit parameters for all 16 intervals are listed in Table \ref{tabpar} and 
the evolution of the parameters with the most interesting variability (\nh, $f$, $\Gamma$, and intrinsic flux) is shown in Figure~\ref{figpars}.
Values for the Fe K ionized wind absorption were also left free, but did not vary significantly over the observations (i.e., they varied by less than the measured error bars).


\begin{figure}
   \plotone{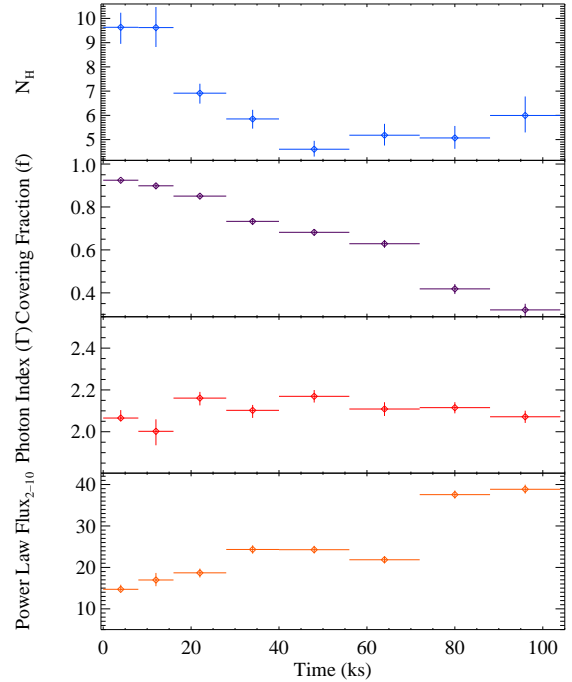}
  \caption{Parameter evolution over the course of observation 3: column density, covering fraction, photon index and unabsorbed 2--10~keV flux.  Note that the decreasing column density and covering fraction seem to have an inverse relationship with the increasing intrinsic flux, while the photon index is relatively stable.  This is unusual behavior for absorption variability may be indicative of a wind absorber scenario. \nh is in units of 10$^{22}$ cm\e{-2} and power law flux is in units of 10\e{-12} \fluxunits.}
  \label{fig3pars}
\end{figure}

\begin{figure}
   \plotone{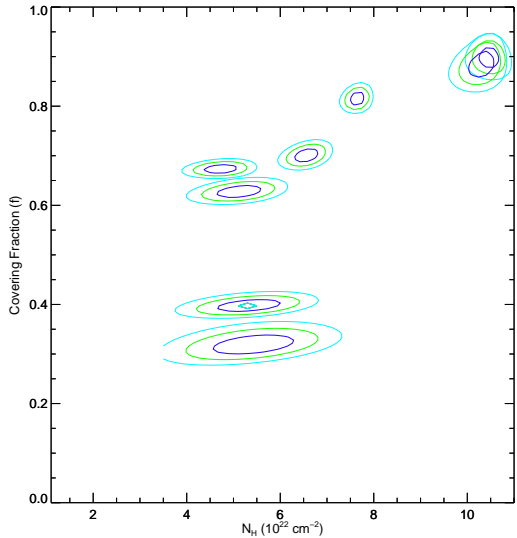}
  \caption{Observation 3 contour plots of \nh versus $f$ for the partial-covering ``patchy'' absorber in the eight sub-intervals, showing that there is very little degeneracy between these parameters.  Levels are at 1$\sigma$, 2$\sigma$, and 3$\sigma$.}
  \label{figcont}
\end{figure}
\begin{figure}
   \plotone{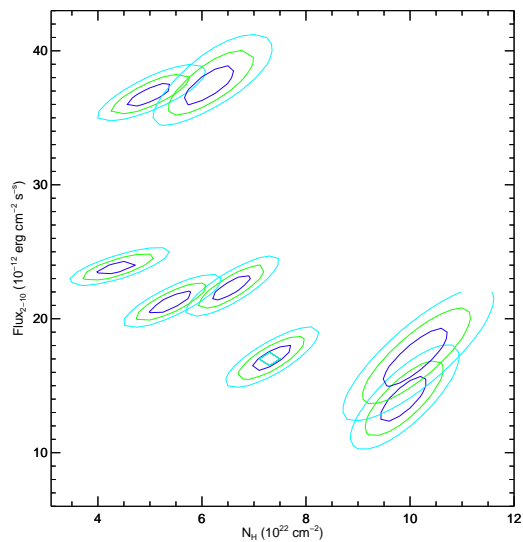}
  \caption{Observation 3 contour plots of intrinsic flux versus \nh for the partial-covering ``patchy'' absorber in the eight sub-intervals.  We see that there is some parameter degeneracy, but that it is not large and it is in the opposite sense to the observed evolution.  Levels are at 1$\sigma$, 2$\sigma$, and 3$\sigma$.}
  \label{figcontff}
\end{figure}

\subsection{The Uncovering of the Source in Observation 3}

In observation 3 we saw a dramatic drop in the absorption by the partial-covering absorber.  
The line of sight covering fraction ($f$) fell from 0.9 to 0.3 over the course of the observation, almost completely uncovering the source.
The column density also dropped significantly (nearly a factor of 2) to 5~\colunits.  Figure~\ref{figpars} shows the rapid decrease in both parameters with 
a corresponding increase in the unabsorbed power law flux.  In order to get a clearer picture of the evolution of these parameters 
we performed additional time-resolved analysis on observation 3, breaking each interval in half and fitting the 8 sub-intervals.
We used the final model as described in section 3.3 with the full covering absorber column density held fixed at the time-averaged value of 1.1~\colunits.

Figure~\ref{fig3pars} shows the evolution of the column density, covering fraction, photon index and intrinsic power law flux for observation 3.
The rapid decrease in both \nh and $f$ could plausibly be due to parameter degeneracy; however the error 
bars seem too small for this to be the case.  A contour plot of \nh versus $f$ for the eight sub-intervals shown in Figure~\ref{figcont} 
verifies that little to no parameter degeneracy is apparent between \nh and $f$.  While there is an expected degeneracy between the column density and flux
of the source shown in Figure~\ref{figcontff}, it is generally small compared to the changes in the parameters and is in the opposite sense to the observed evolution.

The decreasing column density and covering fraction both seem to have an inverse relationship with the increasing intrinsic flux.  
The photon index is relatively stable over the same period. 
This is unusual behavior for absorption variability caused by clumps moving into and out of the line of sight.  
Column density and covering fraction do not typically vary in lock step with one another and we would not expect either parameter to correlate with intrinsic flux.
This may be indicative of an evolution in the absorbing material with luminosity, such as in the wind absorber scenario proposed by Connolley \etal (2014; discussed in Section 4.1). 
We will refer to this layer of absorption (seen primarily in observation 3) as the ``patchy partial-covering'' absorber to distinguish it from the
sometimes partial-covering ``high column density'' absorber which dominates in observations 1 and 4

\subsection{Occultation Event in Observation 4}

\begin{figure}
   \plotone{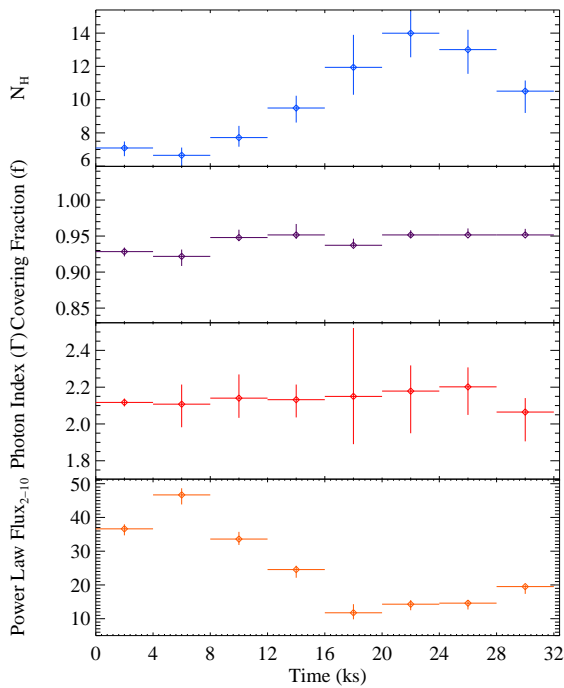}
  \caption{Parameter evolution for the first two intervals (32~ks) of observation 4 divided into eight equal time bins: column density, covering fraction, photon index and unabsorbed 2--10~keV flux.   Here the smooth increase and decrease in column density are reminiscent of occultation events seen in this source and others in the past, though there still seems to be an inverse relationship with the intrinsic flux.  The covering fraction and photon index do not show significant evolution over this time period. \nh is in units of 10$^{22}$ cm\e{-2} and power law flux is in units of 10\e{-12} \fluxunits.}
  \label{fig4pars}
\end{figure}

\begin{figure}
   \plotone{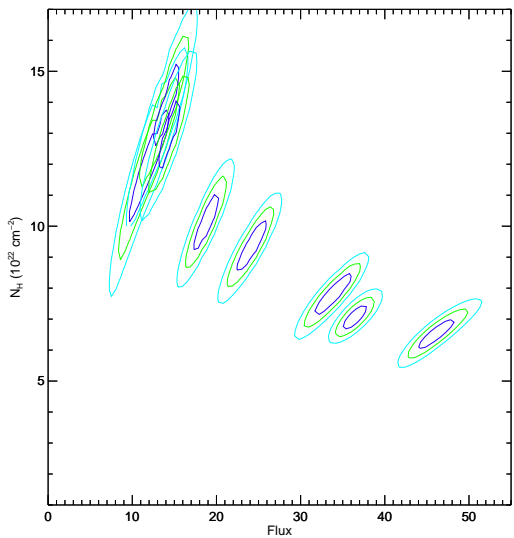}
  \caption{Observation 4 contour plots of intrinsic flux versus \nh for the partial-covering absorber in the eight sub-intervals.  We see that there is some parameter degeneracy, but that it is not large and it is in the opposite sense to the observed evolution.  Levels are at 1$\sigma$, 2$\sigma$, and 3$\sigma$.}
  \label{figcont4}
\end{figure}

Observation 4 shows a rapid increase in column density during the first 32~ks (intervals P13 and P14) with little to no change in the covering fraction.
We broke the first 32~ks down into eight 4~ks sub-intervals and fit just the PN+FPMA data in the 1--40~keV range for computational brevity, using the final model as described in section 3.3.
Parameters for the full-covering absorber, scattered power law, highly ionized wind absorber, and reflection were also held constant.
We find an evolution of the continuum and absorption parameters shown in Figure~\ref{fig4pars}.  
The column density peaked around 20--24~ks, with a seemingly symmetrical profile strongly indicative of a clump of material passing through the line of sight.
Though degeneracy between the column density and intrinsic flux is again present in these fits, the lack of change in the covering fraction makes this behavior very different 
from that seen in observation 3.  Again, the level of degeneracy is less than the observed changes (see Figure~\ref{figcont4}).

\begin{figure}
   \plotone{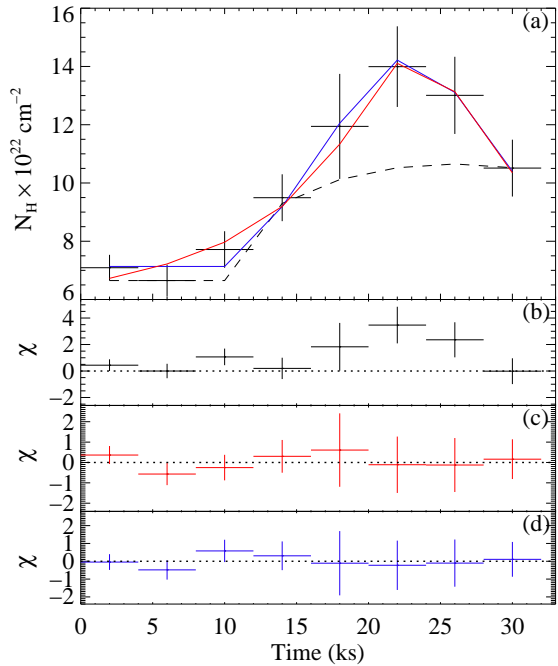}
  \caption{Fitting the occultation event in observation 4 with a solid sphere (dashed black), a beta profile ($\rho \propto r^\beta$; red) and a sphere of linear density profile ($\rho \propto r^{-1}$; blue).  Residuals for the models are shown in (b), (c), and (d), respectively, showing that the latter two models both fit the data reasonably well.}
  \label{fig4nh}
\end{figure}

If this is indeed an occultation event then it is consistent in duration with events seen previously in this source and thought to be due to 
BLR clouds passing through the line of sight (Maiolino \etal 2010).  Using their estimate for $R_{\rm BLR}$ of $\sim \,10^{16}$ cm, and 
assuming a black hole mass of $2 \times 10^{6}$ \Msun we estimate the velocity of the cloud in a circular orbit to be $\sim 1600 km/s$.
Following the analysis of Rivers \etal (2011) for a smooth, symmetrical occultation event assuming a circular orbit around the black hole, 
we fit the absorption profile with a solid sphere, a beta profile ($\rho \propto r^\beta$) and a sphere of linear density profile ($\rho \propto r^{-1}$), shown in Figure~\ref{fig4nh}.
We find a cloud size of $\sim \, 4\times10^{12}$ cm ($\sim 10\,R_{\rm g}$) with a central density of $\sim\,3\times10^{10}$ cm$^{-3}$ for the linearly decreasing density sphere.
These numbers are consistent with physical properties inferred by Maiolino \etal (2010) for the cloud cores, although we do not see evidence of
the same cometary structure inferred in that work (\nh rises again after the first 32~ks of observation 4 and the covering factor is relatively stable over the course of the event).

\subsection{The Additional Soft Component}

From Table \ref{tabpar} we see that the phenomenological Gaussian component at 0.68~keV peaks strongly in observation 3. 
The strength of this feature seems to be anti-correlated with the covering fraction, not showing up at all in observation 1, 
which is fully covered, and only very weakly in observation 4 where the covering fraction is 0.97--1.0.  
This component is therefore clearly not associated with the extended plasma and likely arises within the radius of the variable high column density partial-covering absorber.

Braito \etal (2014) have analyzed the RGS spectrum of observation 3 and found evidence for an increasing ionization in their low column absorber over the course of the observation.
What we are seeing in the MOS/PN may be due to further leakage of the power law below 1~keV due to increasing ionization of the low column full-covering absorber
as the source increases in luminosity. 
The material would have to be quite close to the central source since it is directly correlated with the observed increase in intrinsic power law flux. 
With time resolved fitting on timescales of around 20~ks this would mean the material is at a distance of $\lesssim$10\e{16} cm, 
which is roughly the radius of the BLR and is consistent with being inside the radius of the high column density partial-covering absorber.

Another possibility is the uncovering of actual emission lines such as from O VIII Lyman $\alpha$ at 0.654~keV.  
This could be from highly ionized material very close to the central source, only visible when the full-covering 
low column absorber becomes ionized enough to be semi-transparent at energies below 1~keV.
Braito \etal (2014) also noticed that some broader soft X-ray lines started to emerge during the last part of observation 3, for instance from Mg XI and Mg XII, 
which were much broader than the usual distant narrow line emission, and which seemed to have P Cygni like profiles.
These lines could be associated with a disk wind very close to the central source.
It is worth noting that a visual inspection of the RGS residuals in Figure 4 of Braito \etal (2014) reveals systematically high residuals around 0.65--0.7 keV.


\section{Discussion and Conclusions}

From our spectroscopic analysis of NGC~1365 we begin to understand the full complexity of the multi-layer absorption.
It has long been known that the central nucleus is surrounded by patchy material that provides distant reflection 
and variable absorption of the X-ray corona, both on short timescales (hours) likely associated with eclipsing BLR clouds, 
and on longer timescales (days to months) likely associated with slowly moving material in the torus.  

A full-covering layer of neutral absorbing material with a low column density (\nh $\sim$ 1~\colunits) is required by the data.
This additional ``full-covering low column'' layer is only detectable when the central source is uncovered by the sometimes partial-covering high column absorber.  
Given its stability over at least $\sim$2 months, it is likely not in the inner-most regions of the nucleus.
However, since the extended soft plasma emission is not attenuated by this absorber, it must be closer to the nucleus than the extended plasma.
A third layer of partial covering ``patchy'' neutral absorption is distinguishable from the high column absorber by its unusual variability during observation 3. 
Note that while we only fit a single partial covering absorber component, the drastic difference in behavior during the four different observations leads us to the 
conclusion that we are witnessing multiple layers of absorption dominating at different times.

\begin{figure*}
   \plotone{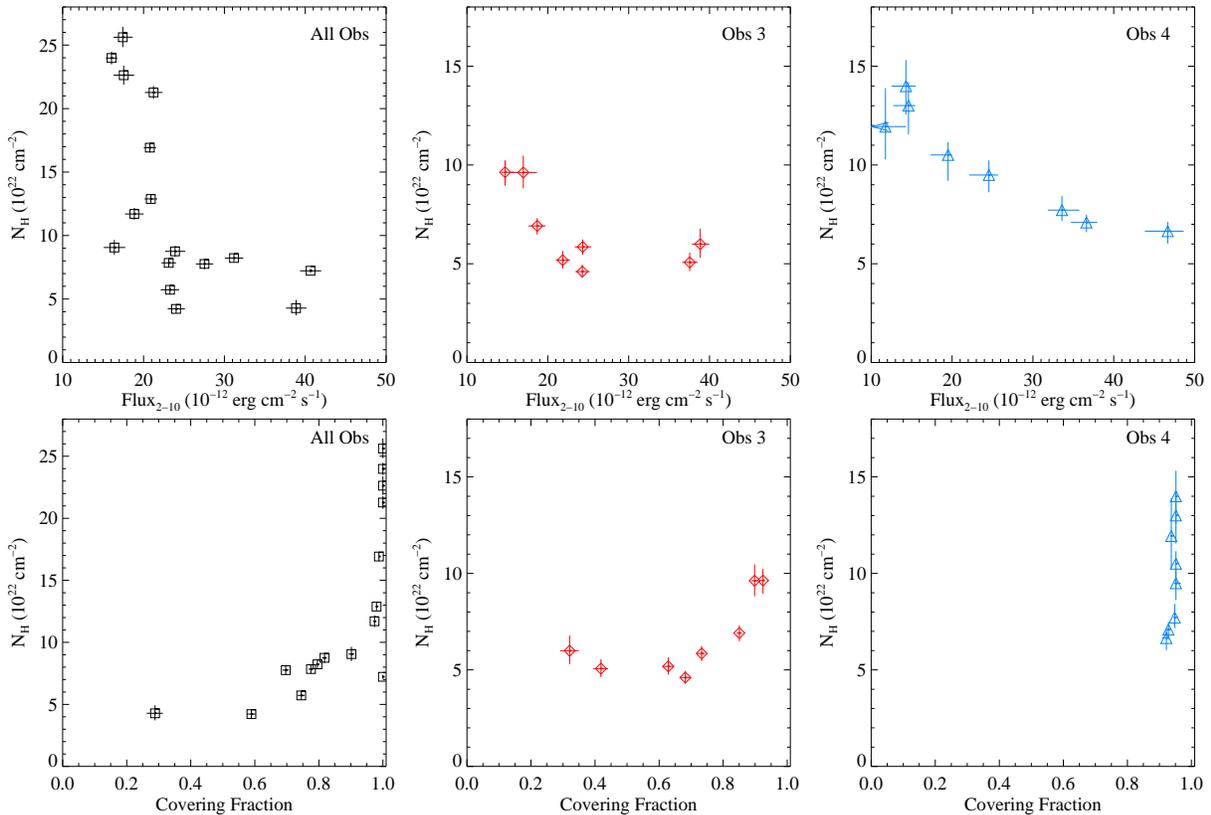}
  \caption{These plots show relationships between spectral fit parameters on the various analyzed intervals. Left: \nh versus intrinsic power law flux and \nh versus covering fraction for the 16 time resolved intervals (P1-16).  Middle:  \nh versus intrinsic power law flux and \nh versus covering fraction for the 8 half-intervals of observation 3.  Right:  \nh versus intrinsic power law flux and \nh versus covering fraction for the 4~ks sub-intervals of the beginning of observation 4.}
  \label{figtrends}
\end{figure*}

\subsection{The Nature of the Variable Absorber}


The uncovering of the continuum by the high column partial covering absorber seen during 2012 December (observation 2) and 2013 January (observation 3) is an unusual event,
particularly the extreme uncovering witnessed in the January observation, where both the column density and covering fraction dropped dramatically.
This is in contrast to earlier observations of fast absorption variability due to ``comet-like'' BLR clouds, which show an increase in the covering fraction as the column density declines,
indicative of a denser core leading a diffuse tail (Maiolino \etal 2010).  
The simultaneous drop in both column density and covering fraction in our third observation 
is inconsistent with this kind of event.

In order to discover the origin of the variability in the absorber we looked for relationships between the parameters that showed the greatest variations.
Figure~\ref{figtrends} shows the relationships between \nh and intrinsic flux, and between \nh and $f$.
The flux plotted here is the unabsorbed power law 2--10~keV flux which corresponds directly to the intrinsic luminosity of the source.
For the 16 intervals, \nh appears to be correlated with flux at the 99.5\% level and with the covering fraction at the 99.9\% level by the Pearson correlation test.
These trends are also seen for the subintervals at around the $\sim$95\% confidence level (although note the small sample sizes).
There is some indication that on longer timescales the photon index softens as the source brightens (correlated at the 99.9\% level for the 16 intervals).
However this correlation is not seen on shorter timescales (for the subintervals of observations 3 and 4).

The relationship between \nh and flux in observation 3 is consistent with the anti-correlation noticed by Connolly \etal (2014) from analyzing {\it Swift} monitoring data of NGC 1365.
Our data do not show a simple linear correlation between these parameters.  It seems clear, however, that when the X-ray source is brighter 
($\gtrsim 2 \times 10^{-11}$~\fluxunits) the column density is much lower ($\lesssim 1 \times 10^{23}$ cm$^{-2}$).  For lower values of 
flux ($\lesssim 2 \times 10^{-11}$~\fluxunits), the column density is much higher ($\gtrsim 1 \times 10^{23}$ cm$^{-2}$).
Connolly \etal (2014) suggested that this could be due to a wind absorber which launches from further out when the source flux goes up.

This patchy wind absorber is only clearly seen in observation 3.  In the other three observations a high column, sometimes partial-covering absorber dominates.
This high column absorber is likely associated with the BLR (as evidence by the occultation event in observation 4 and those seen previously) as well as with the 
torus.  The timescale of the uncovering between observations 2, 3, and 4 is weeks to months rather than hours, indicating that it is either due to a gap in the torus clouds
or to a global attenuation of material.  Since we do not see any other evidence of a temporary drop in overall accretion rate, we favor the former scenario.

One other possibility is that the drop in covering fraction in observation 3 could be explained by a cloud moving out of the line of sight.  
If the cloud were decreasing in size, such as with a drawn out filamentary tail, then instead of the comet-like increase in covering fraction 
we would see a shrinking covering fraction as less and less of the tail covered the source.  This would match our observed absorption parameter evolution,
though it does not explain the correlation between the absorber parameters and intrinsic flux seen in observation 3.  We therefore reject this hypothesis.


\subsection{Size of the X-ray Emitting Region}

In this data set we see changes in the absorbers on multiple timescales.  
We see an initial drop in total column density over the course of $\sim$5 months from 22 to 8~\colunits.
Then in observation 3 \nh goes from 11 to 4~\colunits in $\sim$130~ks while the covering fraction drops to 0.45.
And in the first 20--24~ks of observations 4 \nh rises from 7 to 14~\colunits.
This last is the most rapid change and can be used to place constraints on the size of the X-ray emitting region.
Since the occultation in observation 4 is nearly fully covering, we can infer that the X-ray emitting region is no larger than the size to the occulting cloud: $\lesssim 10\,R_{\rm g}$.

This agrees with previous estimates of the compactness of the X-ray corona from the measured relativistic reflection, 
observed reverberation and BLR occultation events (Walton \etal 2014; Kara \etal 2014; Maiolino \etal 2010).
However it presents a conundrum when we consider the extremely low covering fraction seen in observation 3.
To see such a clear drop in covering fraction would require that the absorber is either very close to the source or made up of clumps/filaments that are smaller than the size of the X-ray emitting region. 
Since the drop in covering fraction occurs slowly over the entire observation and taking into account the low ionization state of the absorbers, 
the latter scenario seems more plausible, possibly due to a patchy disk wind.


\subsection{Summary}

Between July 2012 and February 2013, \nustar and \xmm performed four long-look joint observations of NGC 1365.
We have analyzed the variable absorption seen in these observations in order to characterize the geometry of the absorbing material.
Fortuitously, two of the observations caught the source in an unusually low absorption state, revealing additional complexity which had previously been hidden.
This ``peak between the clouds'' allowed us to see past the typical torus/BLR clouds, which tend to have column densities of around $\sim\,$10$^{23}$ cm$^{-2}$, 
uncovering a patchy absorber with a variable column around $\sim\,10^{22}$ cm$^{-2}$ and a measured covering fraction of $f$=0.3--0.9.
Additionally, we found that this patchy absorber seems to respond to the intrinsic source flux, with the column density and covering fraction dropping as
the source grows brighter.  This could be due to a high luminosity event pushing an absorbing wind out to a large radius where the covering fraction and effective \nh both drop dramatically.  
This latter theory is espoused by Connolly \etal (2014) who noticed an anti-correlation between \nh and luminosity in NGC~1365, a trend which our data confirms.

We also find evidence of an additional constant absorber with a low column density of 1 $\times$ 10$^{22}$ cm$^{-2}$, the geometrical location of which is still unclear.
The ionized wind absorbers seen in this source (Risaliti \etal 2005; Braito \etal 2014) most likely reside closer to the central source than the 
three layers of neutral absorbers we have characterized in this work.

A short occultation event in Feb 2013 was observed, likely due to a BLR clump passing through the line of sight.
We estimate a clump size of $\sim \, 4\times10^{12}$ cm with a central density of $\sim\,3\times10^{10}$ cm$^{-3}$.
From this we also infer a small X-ray corona with a linear dimension of only a few R$_{\rm g}$.

\begin{acknowledgments}

This work was supported under NASA Contract No. NNG08FD60C, and
made use of data from the {\it NuSTAR} mission, a project led by
the California Institute of Technology, managed by the Jet Propulsion
Laboratory, and funded by the National Aeronautics and Space
Administration.  We thank the {\it NuSTAR} Operations, Software and
Calibration teams for support with the execution and analysis of
these observations.  This research has made use of the {\it NuSTAR}
Data Analysis Software (NuSTARDAS) jointly developed by the ASI
Science Data Center (ASDC, Italy) and the California Institute of Technology (USA).
This work has made use of HEASARC online services, supported by NASA/GSFC, and the NASA/IPAC 
Extragalactic Database, operated by JPL/California Institute of Technology under contract with NASA.
This work also made use of data from the \xmm observatory.

\end{acknowledgments}

{\it Facilities:} \facility{NuSTAR}, \facility{XMM}



\end{document}